\documentclass[aps,groupedaddress,superscriptaddress,twocolumn,nofootinbib,prl]{revtex4-1}

\usepackage{graphicx}
\usepackage{color}
\usepackage{amsmath}
\usepackage{amsfonts}
\usepackage{amssymb}
\usepackage{dcolumn}
\usepackage{booktabs}
\usepackage{paralist}
\usepackage[acronym]{glossaries}

\newacronym{WAN}{WAN}{World Airline Network}
\newacronym{ICAO}{ICAO}{International Civil Aviation Organization}
\newacronym{IATA}{IATA}{International Air Transport Association}
\newacronym{SI}{SI}{Supporting Information}
\newacronym{LCC}{LCC}{Largest connected component}

\usepackage{pdfpages}

\begin{document}
\makeglossaries

\title{Revealing the structure of the world airline network}

  \author{T. Verma \footnote{Correspondence and requests for materials should be addressed to T. V. (trivik@ethz.ch)}}
    \affiliation{Computational Physics for Engineering Materials, IfB, ETH Zurich, Wofgang-Pauli-Strasse 27, CH-8093 Zurich, Switzerland}

  \author{N. A. M. Ara\'ujo}
    \affiliation{Computational Physics for Engineering Materials, IfB, ETH Zurich, Wofgang-Pauli-Strasse 27, CH-8093 Zurich, Switzerland}

  \author{H. J. Herrmann}
    \affiliation{Computational Physics for Engineering Materials, IfB, ETH Zurich, Wofgang-Pauli-Strasse 27, CH-8093 Zurich, Switzerland}
    \affiliation{Departamento de F\'isica, Universidade Federal do Cear\'a, 60451-970 Fortaleza, Cear\'a, Brazil}

\begin{abstract}

Resilience of most critical infrastructures against failure of elements that appear insignificant is usually taken for granted. The \gls{WAN} is an infrastructure that reduces the geographical gap between societies, both small and large, and brings forth economic gains. With the extensive use of a publicly maintained data set that contains information about airports and alternative connections between these airports, we empirically reveal that the \gls{WAN} is a redundant and resilient network for long distance air travel, but otherwise breaks down completely due to removal of short and apparently insignificant connections. These short range connections with moderate number of passengers and alternate flights are the connections that keep remote parts of the world accessible. It is surprising, insofar as there exists a highly resilient and strongly connected core consisting of a small fraction of airports (around $2.3\%$) together with an extremely fragile star-like periphery. Yet, in spite of their relevance, more than $90\%$ of the world airports are still interconnected upon removal of this core. With standard and unconventional removal measures we compare both empirical and topological perceptions for the fragmentation of the world. We identify how the \gls{WAN} is organized into different classes of clusters based on the physical proximity of airports and analyze the consequence of this fragmentation. 

\end{abstract}
\maketitle

\section*{Introduction} We seldom hear of a large airspace shutting down. One of the last known examples was triggered by the eruption of the Icelandic volcano, \emph{Eyjafjallaj\"okull}, in $2010$, that led to the cancellation of at least $60\%$ of daily European flights and lasted for five days \cite{Volcano}. Airspace disruptions of this type affect both, global economic activity and daily life of many people. According to the \gls{ICAO}, in 2011, $2.9 \mbox{ billion}$ people used the world airline network to realize business and tourism \cite{PassengerCount}. Understanding the dynamics and resilience of the \glsreset{WAN} to failures is a question of paramount relevance. However, this is not a trivial task. The \gls{WAN} was not planned to be resilient at a global scale in the first place. Instead, it is a network designed to cope up with several economic, political and geographical interests. The resilience of such a network is therefore delicate to quantify and may give us a false sense of global connectivity. 

Algorithms for analysis and design of complex networks have enabled us to quantify complexity and understand the rationale behind their structure and self-organization \cite{Albert2,Barabasi1,Boccaletti,Callaway,Derrible,Opsahl}. Exemplarily, studies on the resilience of various \mbox{infrastructures}, from water transport to the Internet, have provided insights into the evolution and resilience of such networks \cite{Albert1,Brummitt,Kinney,Mamede,Schneider}. In particular, important work has been done in the domain of air transport networks carefully studying the structure \cite{Amaral, Guimera} including extraction of its multilevel modular structure \cite{Sales}. Guimer\`{a} {\it et al.} \cite{Guimera1} analyzed the heterogeneous connection patterns among nodes with different fraction of connections within and outside of their communities that give rise to the dynamics within the air transportation network and other infrastructure networks. In a more recent approach, Cardillo {\it et al.} have studied the emerging features of the \gls{WAN} as a result of the dynamics of cooperation between nodes in the network \cite{Cardillo,Cardillo1}. However, a worldwide view of the \gls{WAN} is not comprehensive without exploring the aftermath of failures in light of such an organization. 

Here, we extract the non-communal hierarchical structure of the \gls{WAN}, analyze this network through a disruption approach and develop weighted measures to understand its fragility. We compare these measures to see how the topology of the network complements the empirical evidence and we reveal a completely new picture of the world airline network. We find that besides the strongly connected core, many important hubs are in fact at the centers of star-like structures which are at the ``periphery", or areas of low economic growth. Upon removing these hubs, the entire star loses its connection to the rest of the network. We show that this mechanism is responsible for the vulnerability of the network and indicate this by identifying different regimes of clustering within the network.

\section*{Results}

Our analysis involves a dataset of the \gls{WAN}, freely accessible at \emph{Openflights} \cite{OpenFlights}, and passengers serviced at each airport during the year $2011$. The \gls{WAN} comprises $N = 3237$ nodes as airports and $L = 18125$ links as direct connections between any pair of airports. An important characterization of a network is its degree distribution which gives us the probability, $P(k)$, of an airport having $k$ connections. The \gls{WAN} data reveals a scale-free behavior with an exponential cutoff, 
\begin{equation}
P(k) \sim k^{-\gamma} exp(-k/k_x),
\end{equation}
with an exponent $\gamma = 1.5 \pm 0.1$ (see \textit{Supporting Information}). The exponential function truncates the distribution around $k_x \approx 180$, which we confirm by analyzing the tail of the distribution in the cumulative complementary degree distribution. In reality, nodes of a physical infrastructure network always depict a truncated distribution as each node can only sustain up to a certain number of connections. 

Each connection might offer multiple alternative flights accounting for its weight. The \gls{IATA} assigns a distinct code to these alternatives. The average weighted degree, i.e. the average number of different flights offered from any airport, of the network is $<k^w> = \sum_{i=1}^{N}{k_i^w}/N = 19.21$. \emph{Frankfurt} airport that falls within the core has the maximum number of flights, $498$, and the maximum number of connections, $255$. Whereas, St. Petersburg airport, Tampa Bay, Florida, a star-like peripheral hub, has only $24$ connections with $24$ flights. The average path length, measured as the average number of minimum connections required from any airport to any other airport, is $<l> = 4.05$. The largest number of connections a passenger needs to travel between any pair of airports is $12$. On average, $33\%$ of the routes can be covered with at most three connections. 

The structure of this network is naturally divided into continents. North America has the largest number of airports followed by Asia and Europe. The classical approach to shed light on the structure of networks is through community detection. We have measured the size of closely formed communities based on the number of flights that exist for a connection between any two airports. To do this, we have used the definition of modularity as introduced by Newman {\it et al.} \cite{Newman2,Newman3} using the algorithm developed by Blondel {\it et al.} \cite{Blondel}. Modularity of a partition is a value that measures the density of links within a community compared to the links that are holding the communities together. This analysis has revealed a total of $20$ well connected communities, identifying economically agglomerated regions of the world, such as the Middle-East, South-East Asia, Alaska and Oceania. Our results are in conformance with the work of Guimer\`{a} {\it et al.} \cite{Guimera} (see \textit{Supporting Information}). This confirms the strong political, geographical and social influence in the development of the air transport network. Nevertheless, the traditional analysis of the network structure fails to uncover the hierarchical structure of the \gls{WAN}. Sales-Pardo {\it et al.} \cite{Sales} have extracted hierarchies within communities for various complex networks, including the \gls{WAN}, wherein the modules of these communities correspond to different levels in the hierarchy. Their approach stems from community detection algorithms that form the basis for identifying cluster of nodes that have more internal links than external. We are interested in a network-wide extraction of hierarchies as a step towards objectively analyzing the vulnerability of the entire system.

We have identified airports and also commercial, cargo and private airstrips that may have been used for a flight recorded by \gls{IATA}. This adds more weight to our analysis as the world is continuously evolving socially and politically and any tie that may have been formed due to a commercial interest may give us insight into the economic connectivity of specific places around the globe. For instance, Alaska has many airports mainly used to serve industries scattered around towns. Our analysis reveals that these airports are completely cut off from the contiguous United States except through major flights to and from Anchorage. This reveals a star-like structure completely different from the country-wide network inside, for example, the United States of America, Canada or Mexico. 

\begin{figure}[ht]
\includegraphics[width = 0.4\textwidth]{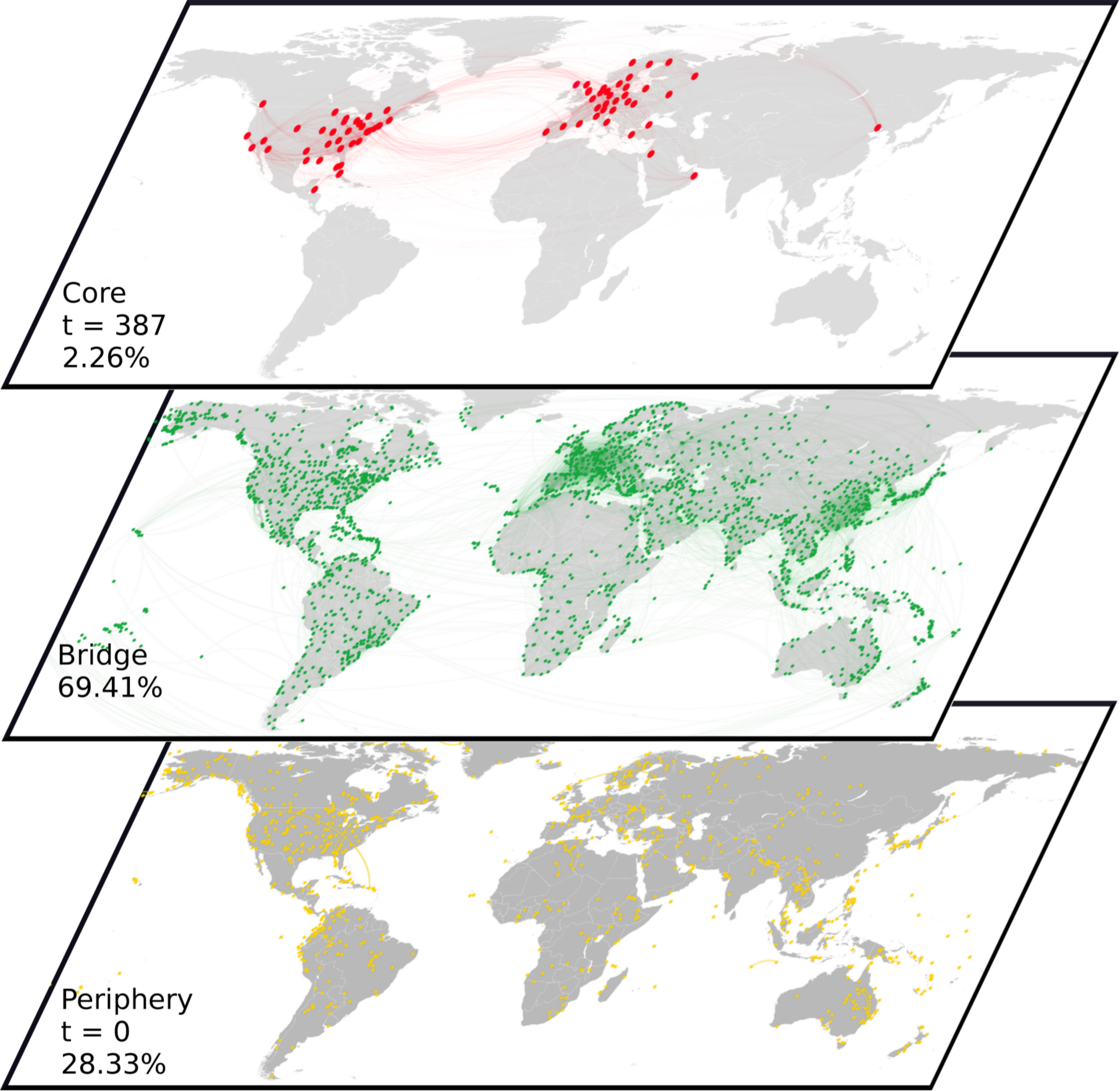}
\caption{
The world airline network divided into three parts. The bottom layer is the \textbf{Periphery} with airports having a zero clustering coefficient. These airports reveal the peripheral world. The top layer is the \textbf{Core} with airports that form the nucleus of the \emph{t-core}, $t = 387$. Nodes in the t-core are part of at least $t$ triangles. This layer shows how well connected some of the major economic hubs of the world are. The intermediate layer is the \textbf{Bridge} with all the remaining airports that act as bridges to connect remote locations to global hubs. \label{fig:WAN}}
\end{figure}

The eclectic mix of star-like structures and a strongly connected core emerges as an interesting global hierarchical structure. In Fig. \ref{fig:WAN} we illustrate three distinct layers of the network - Core, Bridge and Periphery - based on the ``$t-core$ decomposition" of the network. If a connection fails, passengers are typically rerouted through another airport to their destination. This formation is called a \emph{triangle}. To extract the aforementioned structure, we propose a decomposition method based on these triangles. Similar to the $k-core$ method \cite{Dorogovtsev1}, first we remove all nodes that are not part of any triangle. These nodes form the network \emph{periphery} (briefly following the classification of Guimer\`{a} {it et al.} \cite{Guimera1}), as upon removing them a lot of nodes get completely disconnected from the main network. They also form the $0-shell$ and what remains falls within the $1-core$ with nodes being part of at least one triangle. In the next iteration, all nodes with at most one triangle are removed from the network together with their edges and they form the $1-shell$. As we continue to remove nodes that are part of $1,2,3,...t$ triangles, we uncover the bridge of the network with nodes laying at different $t-shells$. Note that removal of a node with $t$ or fewer triangles is done recursively. If removal exposes a new node with now less than or equal to $t$ triangles, it is removed in the current iteration as well. The algorithm stops when each node has been assigned a $t-shell$. The layer that remains at the end is the core of the network (see \textit{Supporting Video}). Each airport in the core is part of at least $387$ triangles. For a list of these airports refer to \textit{Supporting Information}. This indicates that if we remove a connection within the core, there will be numerous other ways to get to the destination. Cohen {\it et al.} \cite{Cohen1} have previously shown that a random scale-free network is vulnerable to intentional attack on hubs and breaks down rapidly. In the \gls{WAN}, this is not the case. Removal of the core, that undeniably consists of many hubs, leads to a minor degradation in the connectivity (the average path length may increase but the world remains connected). In what follows, we focus our attention to the properties of the rest of the network.   

\subsection*{Connectivity}\label{sec:RModel}
Upon removal of the core, most part of the network remains connected and only $8.5\%$ of the airports fall out of the connected cluster. For designing a resilient airline network, analyzing its reaction to catastrophes is significant. We develop a model to understand how the global connectivity is affected due to cancellation of flights or shutdown of airports around the globe. To quantify the loss in connectivity, we measure the fraction of airports that are still part of the largest connected component, $S(q)$, and observe it as a function of 
\begin{inparaenum}[\itshape a\upshape)]
\item the fraction of airports being shut and
\item the fraction of connections getting canceled. 
\end{inparaenum}
Note, that in our case $S(0) = 1$, as we start with merely one connected cluster.

\begin{figure}[ht]
\includegraphics[width = 0.4\textwidth]{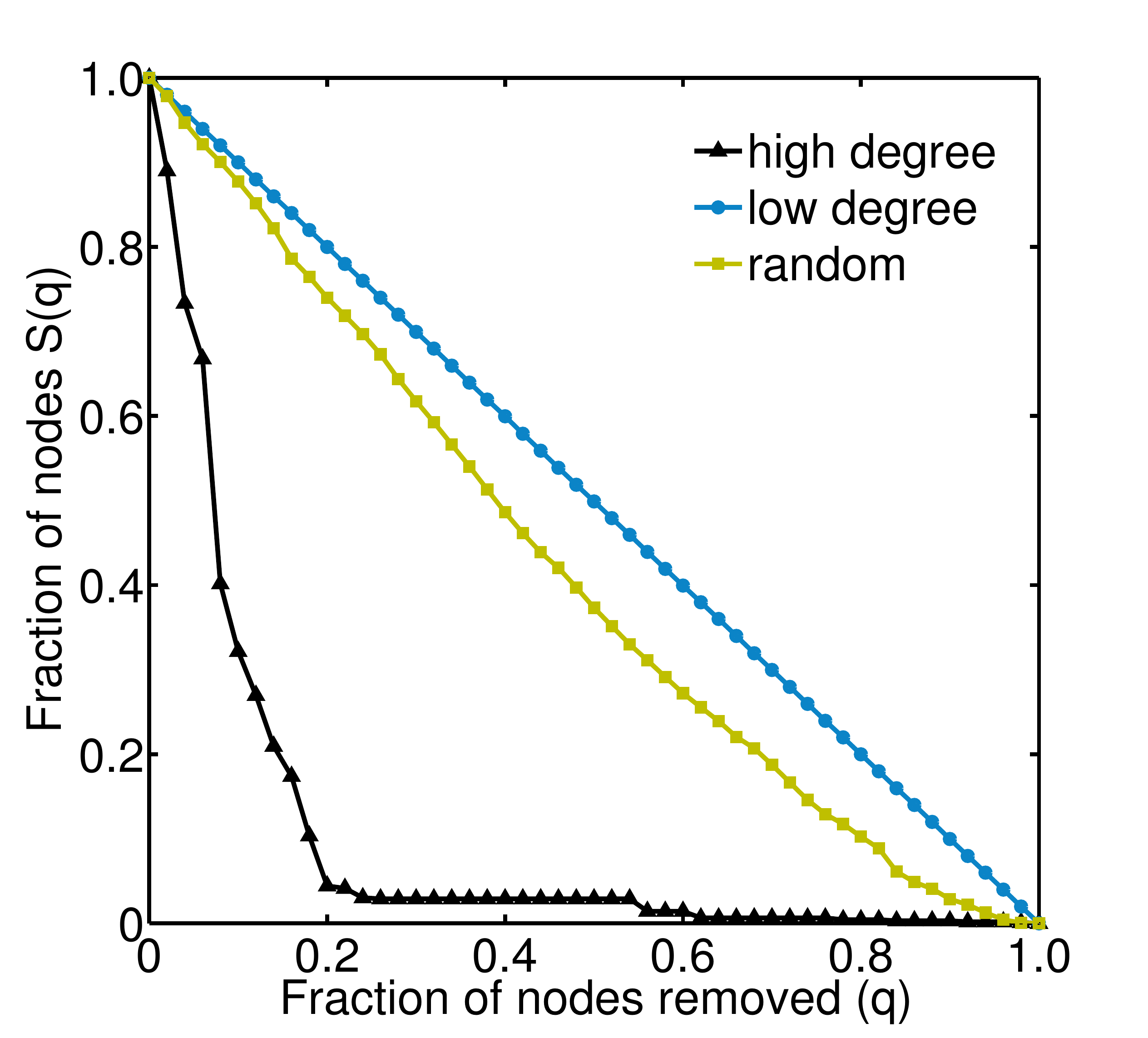}
\caption{
Drop in the size of the largest connected cluster of the \gls{WAN} against removal of \textbf{nodes}. \emph{High degree} refers to a conventional targeted removal strategy wherein each subsequent step corresponds to removing upto a fraction, $q$, of nodes with the highest degree. In \emph{low degree} strategy, each subsequent step corresponds to removing upto a fraction, $q$, of nodes with the lowest degree. The random removal strategy is an average over $500$ statistically independent simulations. \label{fig:NodePercolation}}
\end{figure}

We analyze the connectivity of the network by sequentially shutting down airports using two different strategies. In the first one, at each step, we remove airports with the highest degree. Figure \ref{fig:NodePercolation} shows that upon removal of the highest degree airports, the network rapidly disintegrates into many small clusters and the size of the largest connected component drops significantly. When the most connected airports are not functional anymore, the long haul flights that give the network a small-world characteristic \cite{Guimera} also break down, explaining the sudden drop in connectivity. These airports naturally fall in the core of the network. By contrast, upon removal of the lowest degree airports, the size of the largest cluster decays linearly. As shown in Fig. \ref{fig:WAN}, this network is hierarchically structured with a well connected core and a tree-like structure at the periphery. No removal of low degree airports can affect the rest of the network because they are in the periphery. Thus, the core holds the network together while the leaves of the tree are pruned. The aforementioned analysis strengthens our argument of airports with no clustering being peripheral in the sense that they are at the extremes of the network. 

Airports shut down only in extreme cases. Subsequently, we focus on the most common scenario where connections are canceled. As we will show, the picture is significantly different when connections fail instead of airports. 

\subsection*{Passenger Flux}\label{sec:PModel}
In an effort to remove connections from the network, we need to define the relevance of the connections. Passenger flux seems ideal to describe the relevance of a connection. Data is available for the number of passengers being serviced at each airport in $2011$ and we define the relevance of a connection as follows,
\begin{equation}
p_{ij} = p_i \frac{k^w_j}{\sum_{k \in neigh(i)}{k^w_k}},
\label{eq:flux}
\end{equation}
where $p_{ij}$ is the passenger flux on the connection from $i$ to $j$. $neigh(i)$ is the set of all nodes directly reachable from $i$. $k^w_i$ gives the total number of flights available from airport $i$ and $p_i$ is the number of passengers at airport $i$. 

This division may overestimate the flux of passengers on a connection but keeps its relevance intact as the number of passengers taking a connection depends on further available connections from the destination airport. In addition, it is also a simple measure to identify destinations that are popular tourism and business spots. This method does not explicitly take into account the frequency of flights per day between two airports, but that has implicitly been accounted for by considering the number of passengers using that connection. We use the above measure following a basic ideology of the airline network, i.e. a flight is only as important as the number of passengers using it. This provides us with an empirical measure on the network. We constitute another measure based on connectivity to complement the empirical evidence from the data.

\subsection*{Degree of Connectivity}\label{sec:CModel}
The degree of connectivity, $c_{ij}$, characterizes each connection using the topological information about the network. The topological clustering coefficient \cite{Watts1} of each node in the network which measures the degree to which neighbors of a node cluster together is defined as follows,
\begin{equation}
C_c(i) = \frac{2E}{k_i(k_i - 1)},
\end{equation}  
where $k_i$ is the number of direct connections of airport $i$ and $E$, the number of connections that exist between the first neighbors of $i$. Each connection has an integer weight $w_{ij}$ given by the number of flights. We define degree of connectivity as follows,
\begin{equation}
c_{ij} = C_c(i) w_{ij},
\end{equation}
which is asymmetric in nature, i.e. $c_{ij}$ is not necessarily equal to $c_{ji}$. Not all passengers fly back to the source airport or take the same connections on their return journey. If the clustering coefficient is very low, then a \emph{connection} from this airport with very few flights will be extremely important due to very few alternate paths that passengers can take to their destination. No passenger \gls{WAN}ts to take five connections to a destination that is at a relatively short distance. Thence, we focus more on a removal strategy based on low degree of connectivity and show how seemingly irrelevant details of a complex network might add to its vulnerability. For an elaborate justification of this concept, we have included an example in \textit{Supporting Information}. 

Connections that serve the least number of passengers or have the lowest degree of connectivity are most often overlooked and likely to be in the periphery. An important long-range connection cannot affect the connectivity of the network because of a high degree of redundancy, for example, cancellation of the longest flight, which is between Newark Liberty International airport in New Jersey, USA and Changi Airport, Singapore, will not restrict the mobility of passengers as more than fifteen one-stop alternatives exist for this route. But many short-range connections within the periphery, such as from Anchorage, Alaska, to Honolulu, Hawaii (an important tourist hub), have only one option and such cancellations can result in stranded passengers since very few passengers would accept to take even two connections that take a much longer detour (about twice the flying time of the direct connection).

\begin{figure*}[ht]
\includegraphics[width = 0.9\textwidth]{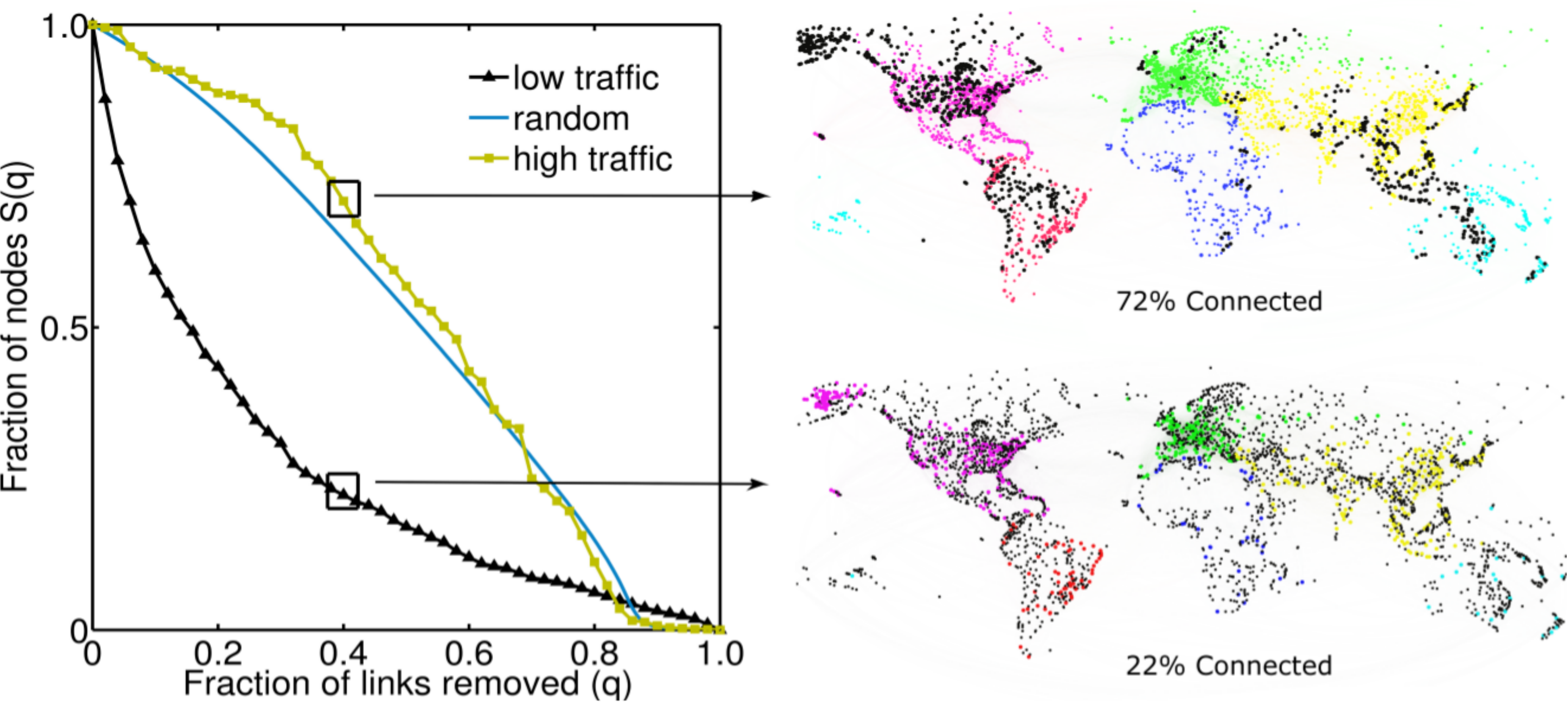}
\caption{
Drop in the size of the largest cluster of the \gls{WAN} against removal of links. Connections are ranked according to the number of passengers using it. In \emph{high traffic} removal, each subsequent step corresponds to removal of all connections up to a fraction, $q$, with the highest number of passengers. In \emph{low traffic} removal, each subsequent step corresponds to removal of all connections up to a fraction, $q$, with the lowest number of passengers. The random removal strategy is an average over $500$ statistically independent simulations and each step removes a fraction, $q$, of connections chosen at random. After removing $40\%$ of the busy connections, $72\%$ of the network is still connected, shown in the top-right map. The bottom right map shows that after removing the same fraction of idle connections, the world disintegrates completely, revealing the vulnerable nature of the periphery of the network ($22\%$ connected). The black nodes are not part of the largest connected cluster. The remaining colors represent different continents and show the nodes that are part of the largest connected cluster. \label{fig:ResilienceOther}}
\end{figure*}

Following the above arguments and defining relevance of a connection based on empirical evidence and topological information, we remove a certain fraction of connections. Here, relevance of a connection directly translates into traffic or connectivity. A \emph{high traffic} removal would affect the busiest connections. Whereas, a \emph{low traffic} removal would focus on idle connections. Figure \ref{fig:ResilienceOther} shows that the world fragments into different parts upon removing a small fraction ($20 \%$) of idle connections while it is almost fully connected upon removing busy connections, using passenger flux as relevance. During an economic crisis or loss suffered by an airline company, the typical flights to be canceled are the ones that carry the smallest number of passengers. Cancellation of such flights, however, causes more damage to the global connectivity than flights that connect hubs. The analysis using \emph{degree of connectivity} shows similar results. Moreover, the effect of an idle removal is magnified using passenger flux over degree of connectivity. Both these observations give enough topological and empirical evidence for the importance of the peripheral connections in the \gls{WAN}. The periphery is huge and consists of many airports and connections between these airports. To understand what kind of connections lay in the periphery, we analyze the resilience using the most basic information of the connections - the number of flights associated with it.

\subsection*{Flight Model}\label{sec:FModel}
Airports might offer more than one flight for a direct connection between them. We define for each connection an integer weight, $w_{ij}$, given by the number of alternate flights that are available on that connection. The data set does not give us precise information about the frequency of flights. Each alternate flight has a different \gls{IATA} code, i.e. they are operated by either different companies or at different times. Passengers have the option to choose among the flights from source to destination based on their preferences (price, time of flight, reliability and quality-of-service, etc).

We study the connectivity of the network upon removing connections following the rank of their weight. We found that the \gls{WAN} is quite resilient to breakdowns in \emph{frequent} connections with multiple flights as it has a high degree of redundancy with many paths between any pair of airports with high degree (see \textit{Supporting Information}). However, the global connectivity of the network, upon removing  \emph{rare} connections with the least number of flights, is lower than any other removal strategy. Our common sense would mislead us to believe that disruption of peripheral connections would not lead to loss in global connectivity and therefore predestined to face economic cuts in case of a crisis. However, our result shows that the periphery is weakly connected in terms of possibles routes as well as the number of different flights between the same pair of airports and could render a large part of the world inaccessible. Refer to \textit{Supporting Information} for a visual representation of how global connectivity evolves with the sequence of failures. In order to explain this phenomenon, we study the clustering properties of the \gls{WAN} in depth.

\subsection*{Clustering}\label{sec:Clustering}
Among other reasons airports form connections based on physical proximity. Airports can be clustered differently. We have identified three regimes of clustering using the physical length of connections.

Firstly, we define the weighted clustering coefficient, $C_w(i)$, as defined by Barrat {\it et al.} \cite{Barrat}
\begin{equation}
C_w(i) = \frac{1}{s_i(k_i - 1)}\sum_{j,h}\frac{(1/d_{ij} + 1/d_{ih})}{2}a_{ij}a_{ih}a_{jh},
\end{equation} 
which is a measure of local cohesiveness of neighbors of an airport that takes into account the intensity of the connections given by its euclidean distance, $d_{ij}$,  between airports $i$ and $j$. $s_i$ is the strength of an airport $i$ defined as $\sum_{j \in neigh(i)}{1/d_{ij}}$. $k_i$ is the number of connections from an airport (out-degree). Lastly, $a_{ij}$ is either $0$ or $1$ depending on the absence or presence of a connection between airports $i$ and $j$, respectively. In this way we consider the total relative weight of the closed triplets of any airport with respect to the strength of the airport. The topological clustering coefficient, $C_c$, is obtained by simply marking $d_{ij} = 1$ for all connections.

The following regimes of nodes can be distinguished,
\begin{enumerate}
\item Peripheral: Nodes that have no connections between their neighbors, $C_c(i) = C_w(i) = 0$. 
\item If $C_c(i) \geq C_w(i)$, the interconnected triplets are formed by connections with large distances and hence are \emph{global}. 
\item If $C_w(i) > C_c(i)$, the interconnected triplets are formed by connections with short distances and hence are \emph{local}.
\end{enumerate}

Most airports in the first category have really few connections but some of them are busy airports having a significant number of connections to and from cut-off regions, and hence we call them peripheral hubs. For instance, St. Petersburg Airport, Tampa Bay, Florida, is a huge tourist destination consisting of $24$ connections and no clustering. In the remaining categories, nodes are distributed equally with low degree airports having a low number of passengers and high degree airports having a higher number of passengers. Examples of airports in the local category include Frankfurt and JFK, New York. Atlanta and Domodedovo, Moscow, fall in the global category. Each continent has at least $40\%$ airports with local clustering pointing to a continental evolution. When the continents are connected to form a world airline network, there is no significant change in the fraction of nodes that form local clusters, indicating that the airline networks typically evolved at the level of continents.

\begin{figure*}[ht]
\includegraphics[width = 0.9\textwidth]{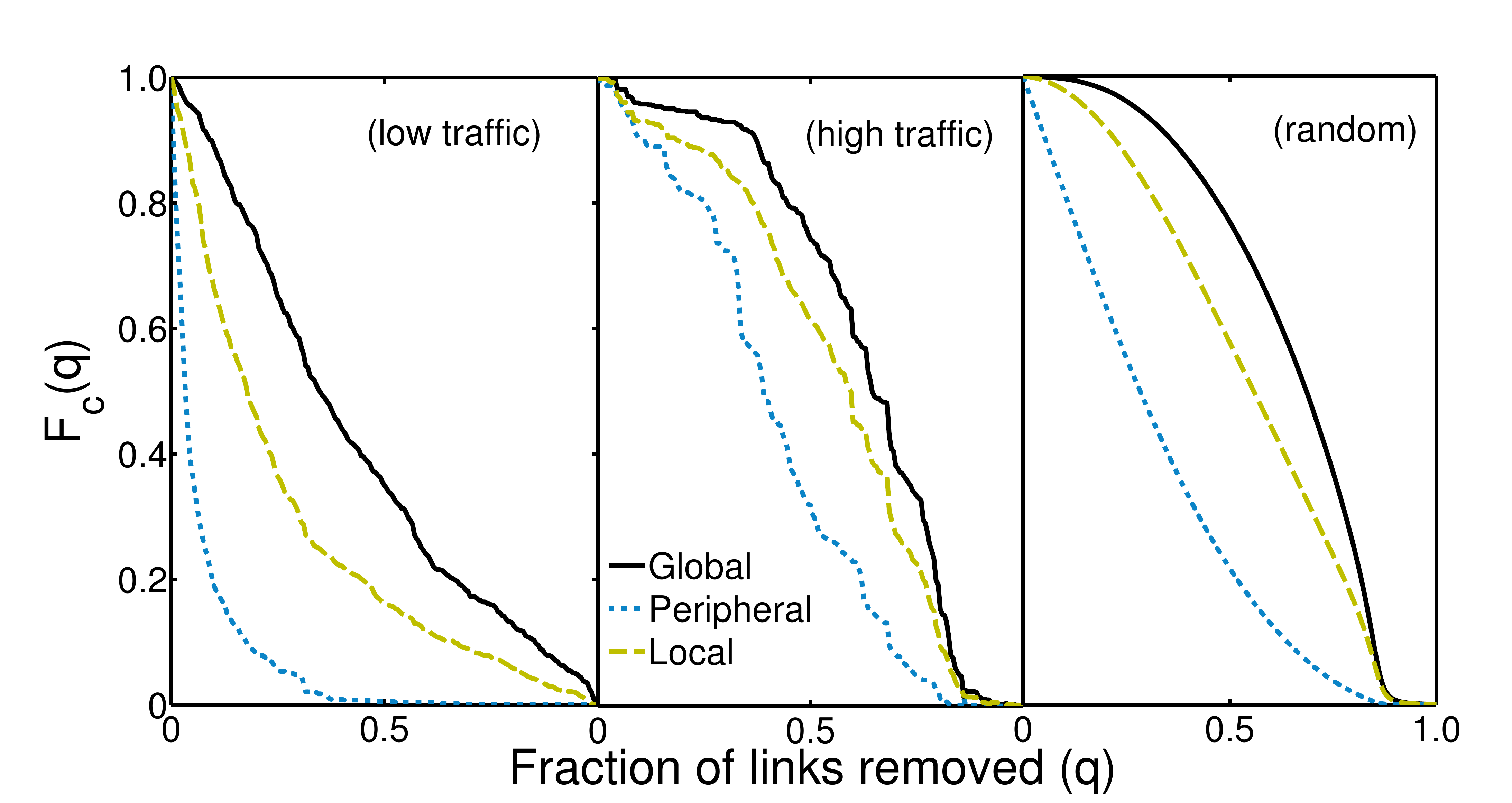}
\caption{
Drop in fraction of airports within a cluster regime. $F_c(q)$ is the fraction of airports belonging to a cluster regime $c$ after removal of a fraction of connections $q$. The first frame shows a drop in airports that belong to the largest connected component based on an \emph{low traffic} removal. The subsequent frames show the same for \emph{high traffic} removal and an average over $500$ random removals. In all cases, peripheral hubs ($C_c = C_w = 0$) drop out of the largest connected component first. \label{fig:CCComparison}}
\end{figure*}

In almost all cases of removal scenarios, airports with zero clustering are the ones that are mainly responsible for a drop in the resilience of the network, followed by locally clustered airports and then the global ones. For instance, Fig. \ref{fig:CCComparison} shows the drop in fraction of airports within each cluster regime upon removal of connections. Here, link relevance is defined using passenger flux. All airports that have a zero clustering coefficient are disconnected from the largest connected component first. Since these airports constitute the peripheral hubs of the network, they take down the extremities of the network with them (Fig. \ref{fig:WAN} shows a large fraction of airports situated in relatively inaccessible areas of the world). This explains a sudden drop in the connectivity of the network. 

\section{Discussion}\label{sec:Discussion}
In summary, we found that even though the network has a core resilient structure, which guarantees intercontinental connections, most of the world is accessible through peripheral connections. Clustering plays a crucial role as most airports that have no alternative connections to their destinations are the ones that make the \gls{WAN} most vulnerable. This is surprising since the traditional strategy of studying the community structure (see \textit{Supporting Information}) and targeted removal scenarios gives no evidence of such a hierarchical ordering in the network. The global connections are economically and politically significant with the world moving towards a free society for travel and living, but the local connections serve the population of the region in extending tourism and business. 

A possible reason for the existence of a strongly connected core and a weakly tied periphery is the necessity of airline companies to cope with the minimization of flying time and the maximization of profit. A flight is only profitable if there is a minimum number of passengers per flight. Flying time is ideally minimized with a fully connected network with every pair of airports directly connected. Yet, this is only economically reasonable between highly populated areas or regions of intensive economic activity, which are served by the strongly connected core. The scanty number of passengers traveling to and from remote regions only justifies the creation of a star-like network, with a peripheral hub in the center, as we found in the periphery of the \gls{WAN}.

Future work could account for temporal evolution of the resilience of the network. With an improved model with accurate passenger count on each connection and the frequency of flights, we can extract the community structure based on influx of passengers to obtain an even more realistic assessment of resilience. It would also be interesting to study passenger flows in different locations of the world and reveal the travel patterns toward which the current generation is moving. This could be beneficial for airline companies not only to maximize their profit but also to redesign the network to make it more resilient. 



\section{Methods}

\textit{\gls{WAN}:} The flight data (airports, airlines, routes and geo-locations) was obtained from \emph{OpenFlights} \cite{OpenFlights} as of May, 2013. This data contains some \emph{circular} connections, i.e. a flight may go from $A$ to $B$ and not return directly to $A$. Instead, this flight follows a path from $A$ to $B$ to $C$ and back to $A$. To simplify our analysis, we have made the adjacency matrices symmetric by replicating each unidirectional connection in the opposite direction. This is justified by the fact that only very small airline companies have circular connections and merely in remote parts of the world. 

\textit{Passengers:} The passenger data was obtained from \emph{The World Bank} dataset sourced through Civil Aviation Statistics of the World and \gls{ICAO} staff estimates for the year 2011. This data has passenger count for every country that is registered with scheduled air carriers of that country. Changes in air transport regulations in Europe have made it more difficult to classify traffic as scheduled or unscheduled. Thus recent increases shown for some European countries may be due to changes in the classification of air traffic rather than actual growth. We have divided the passenger data among airports of the countries based on their relative weighted degree using the following equation,
\begin{equation}
p_i = p_c \frac{k_i^w}{\sum_{j \in c}{k_j^w}},
\end{equation}
where $p_i$ gives the number of passengers serviced at any airport $i$ from country $c$, $p_c$ is the passenger data of the country $c$ and $k_i^w$ is the number of alternative connections from any airport $i$, i.e. the weighted degree. This approach gives us an indication of the relevance of connections originating from each airport (see \textit{Supporting Information}).

\newpage

\begin{acknowledgments}

We acknowledge financial support from the ETH Risk Center and European
Research Council through Grant FlowCSS No. FP7-319968. We also thank Vitor Hugo Louzada for motivating discussions and valuable comments.

\end{acknowledgments}

\section*{Authors Contributions}
T.V., N.A.M.A., and H.J.H. wrote the main text, prepared the simulations, and discussed the results.

\section*{Additional Information}
\textbf{Competing financial interests:} The authors declare no competing financial interests.

\clearpage


\section*{Supplementary material}

# Supporting Information: Revealing the structure of the world airline network

Trivik Verma,[1, *] Nuno A. M. Araújo,[1, †] and H. J. Herrmann[1, 2, ‡]

[1]*ETH Zürich, Computational Physics for Engineering Materials,
Institute for Building Materials, Wolfgang-Pauli-Strasse 27, HIT, CH-8093 Zürich, Switzerland*
[2]*Departamento de Física, Universidade Federal do Ceará,
Campus do Pici, 60455-760 Fortaleza, Ceará, (Brazil)*

## S1. FLIGHT MODEL

Figure S1 visually depicts how the world fragments into different parts as the peripheral connections with the least number of flights are removed. Frame 2 of this figure illustrates that 45% of airports fall out of the largest connected component upon removing connections with merely one flight ($\approx$ 60% connections). In frame three of the figure, it is clear how only the global hubs are connected when connections carrying up to *eight* flights ($\approx$ 96% connections) have stopped plying. These flights are connecting small fishing industries in remote islands or making chartered connections to sectors outside of economic and political hubs. These also include connections to scarcely populated areas in Siberia, Alaska, Northern Canada, Papua New Guinea, the Sahara Desert, etc.

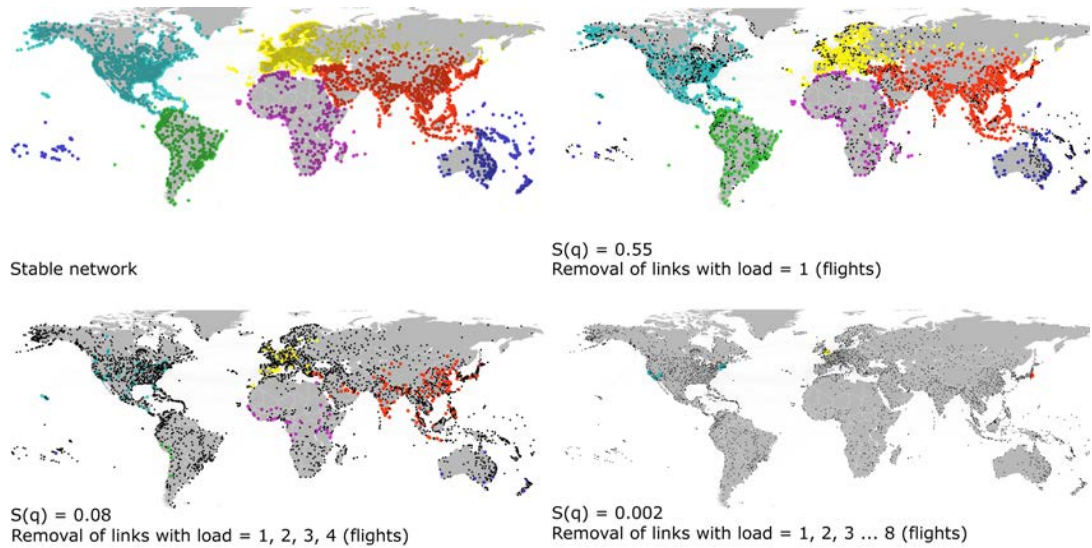

FIG. S1. Subsequent stages of WAN under a *rare* removal of links as ranked by number of *flights*. The colors represent different continents but also only the largest connected component of the network. Black nodes are airports that have been disconnected from the largest connected component but still may be functioning (not completely isolated) within their own small clusters.

## S2. RATIONALE BEHIND *DEGREE OF CONNECTIVITY*

Figure S2(a) shows node $b$ has zero clustering. Link $b - a$ is extremely important as it is the only connection between $b$ and $a$. Whereas, in fig. S2(b), $a$ can be reached from $b$ in many ways and link $b - a$ is redundant in a network sense.

---

[*] trivik@ethz.ch
[†] nuno@ethz.ch
[‡] hjherrmann@ethz.ch

<a>a</a>

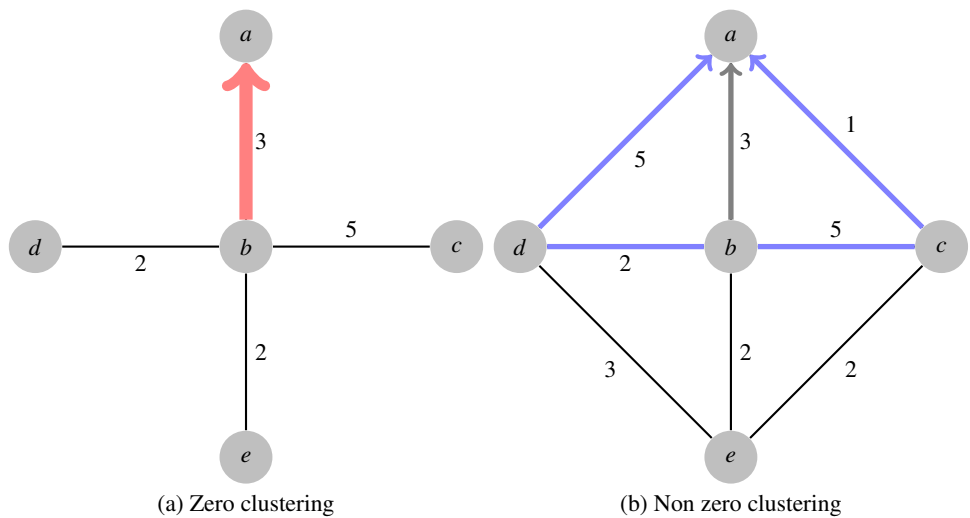

FIG. S2. Illustration of the rationale behind the *degree of connectivity*. Each link is characterized by the number of flights on that connection. In (a), the clustering coefficient of $b$ is zero and hence $b-a$ is a very important link as there is no other way to reach $a$ via $b$. However, in (b), the clustering coefficient of $b$ is greater than zero, i.e., nodes in the neighborhood of $b$ are also connected to each other. This implies that the a higher *degree of connectivity* gives a lower "importance" to the link and also clarifies our rationale behind unconventional removal scenarios.

### S3. t-CORE DECOMPOSITION

Figure S3 shows the core of the World Airline Network. It comprises 73 airports that are extremely well-connected and provide a strong foundation for air travel.

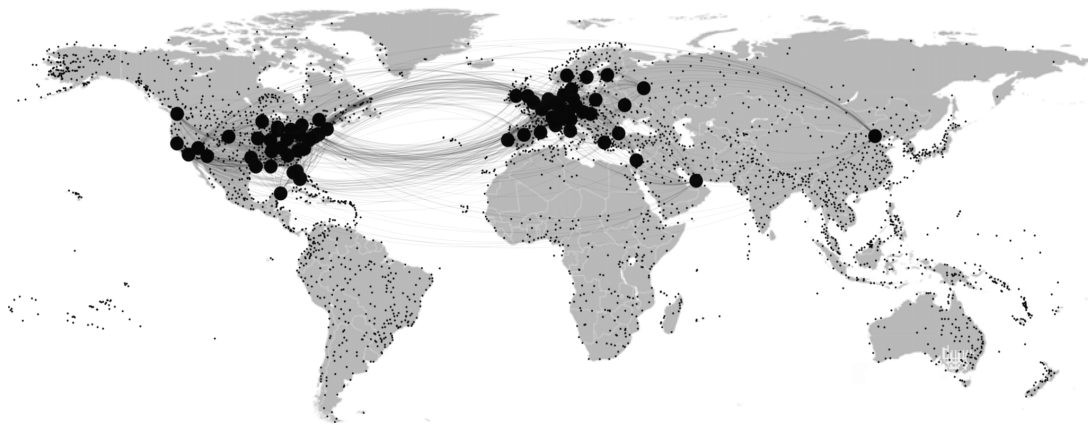

FIG. S3. Seventy three airports that belong to the core of the network. All these airports fall in the t-core with $t = 387$.

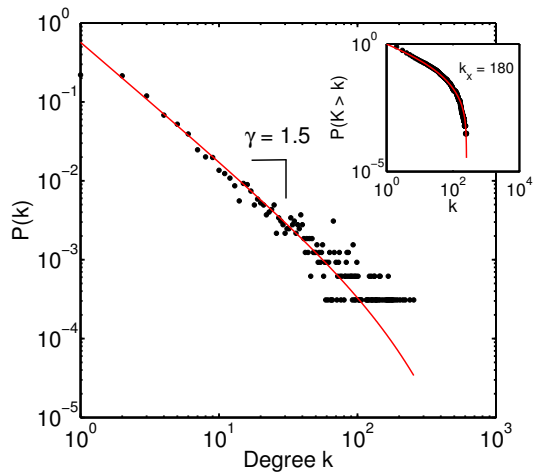

FIG. S4. Degree distribution of WAN plotted in a log-log scale. The degree is a measure of the connections of an airport and follows a scale-free behavior with an exponential cut-off, $P(k) \sim k^{-\gamma} exp(-k/k_x)$ with an exponent $\gamma = 1.5 \pm 0.1$ and truncates at $k_x = 180 \pm 5$. The **inset** shows the cumulative complementary distribution for the network in a log-log scale with the same exponents as above. There is an exponential decay for the number of airports with large degree.

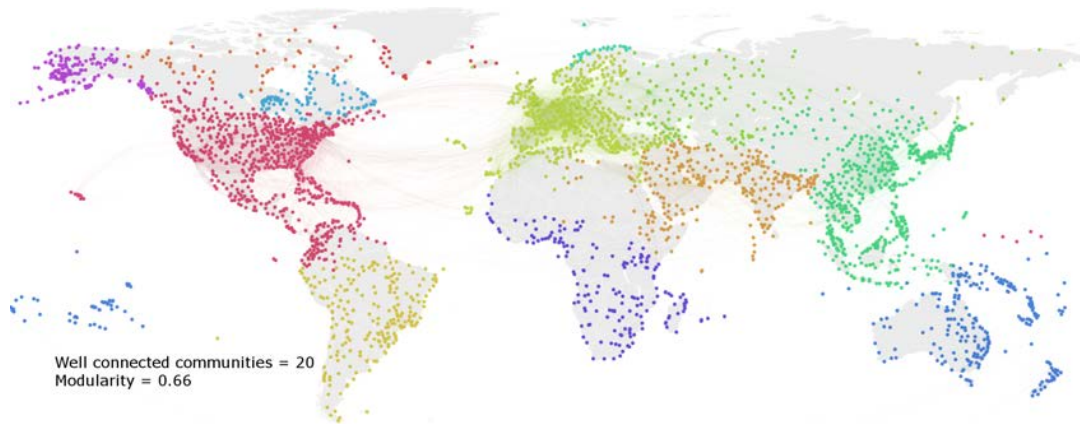

FIG. S5. Closely held communities existing in WAN based on the number of alternate flights between each pair of airports. Nodes are locations of the airports and colors represent different communities. A high value of modularity ($M \to 1$) indicates presence of well-formed communities.

4

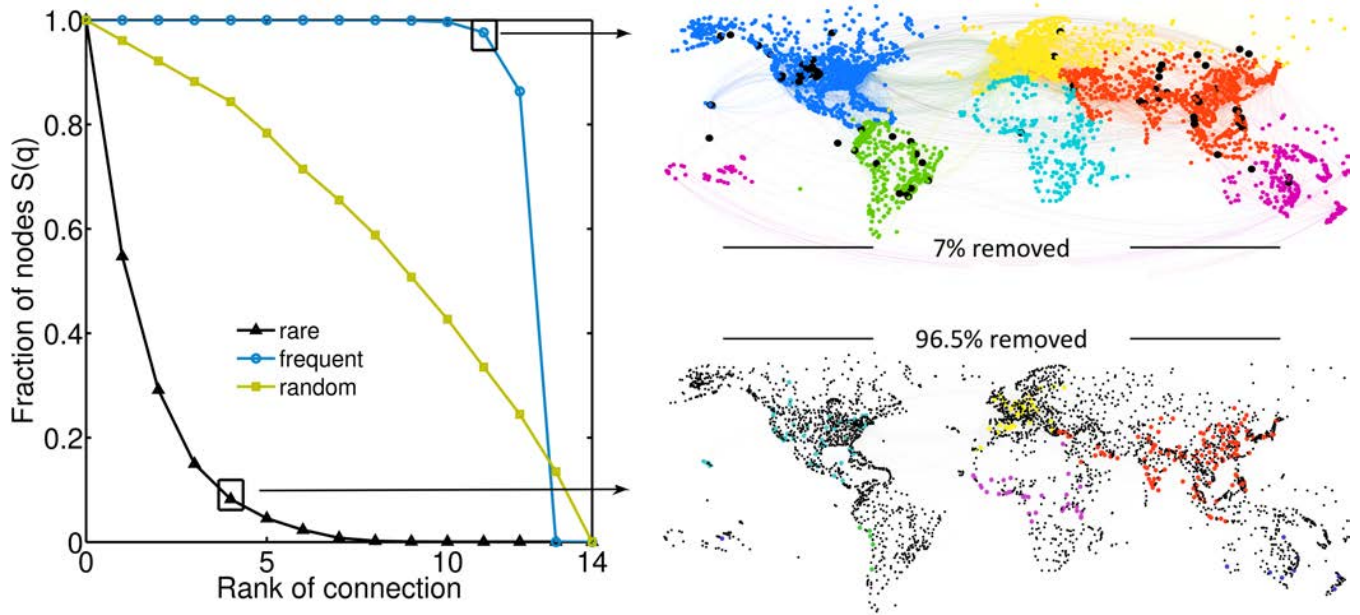

FIG. S6. Drop in the size of the largest connected cluster of WAN against removal of links. Connections are ranked according to their weight given by the number of alternative flights. In *frequent* removal, each subsequent step corresponds to removal of all connections with the highest weight. In *rare* removal, each subsequent step corresponds to removal of all connections with the lowest weight. The random removal strategy is an average over 500 statistically independent simulations and each step removes all connections with a random probability. After removing connections that offer at least four alternatives (7.5% connections), a large part of the network is still connected, shown in the top-right corner. The bottom right map shows that after removing connections that offer at most four alternatives (96.5% connections), the world disintegrates completely, revealing the vulnerable nature of the periphery of the network. The black nodes are not part of the largest connected cluster. The remaining colors represent different continents and show the nodes that are part of the largest connected cluster.

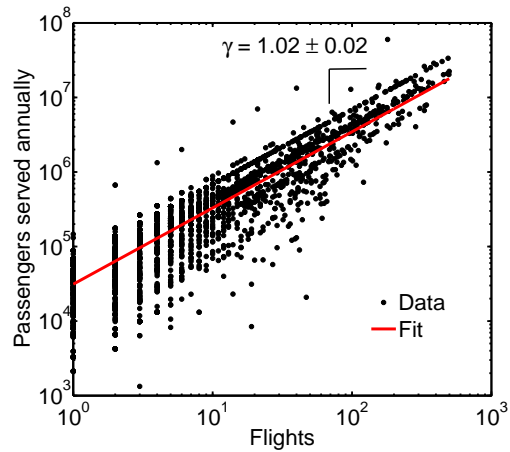

FIG. S7. Scatter plot of the number of passengers served at each airport annually versus the number of available alternative connections at each airport. The plot is in a log-log scale. Several airports with the same number of connections has varied passenger count. We describe this behavior by adding the red line to the curve with the following scale-free exponent, $\gamma = 1.02 \pm 0.02$.




\begin{thebibliography}{10}

\bibitem{Volcano}
America, V.
\newblock \url{http://www.voanews.com/}.

\bibitem{PassengerCount}
\gls{ICAO}.
\newblock \url{http://goo.gl/LCgb5f}.

\bibitem{Albert2}
Albert, R., Jeong, H., and Barab{\'a}si, A.-L.
\newblock {\em Nature}{ \bf 406}, 378--382 (2000).

\bibitem{Barabasi1}
Barab{\'a}si, A.-L., Albert, R., and Jeong, H.
\newblock {\em Physica A}{ \bf 281}, 69--77 (2000).

\bibitem{Boccaletti}
Boccaletti, S., Latora, V., Moreno, Y., Chavez, M., and Hwang, D.-U.
\newblock {\em Physics Reports}{ \bf 424}, 175--308 (2006).

\bibitem{Callaway}
Callaway, D.~S., Newman, M.~E., Strogatz, S.~H., and Watts, D.~J.
\newblock {\em Phys Rev Let}{ \bf 85}, 5468--5471 (2000).

\bibitem{Derrible}
Derrible, S.
\newblock {\em PloS One}{ \bf 7}, e40575 (2012).

\bibitem{Opsahl}
Opsahl, T. and Panzarasa, P.
\newblock {\em Social Networks}{ \bf 31}, 155--163 (2009).

\bibitem{Albert1}
Albert, R., Albert, I., and Nakarado, G.~L.
\newblock {\em Physical Review E}{ \bf 69}, 025103 (2004).

\bibitem{Brummitt}
Brummitt, C.~D., D'Souza, R.~M., and Leicht, E.
\newblock {\em Proc Natl Acad Sci USA}{ \bf 109}, E680--E689 (2012).

\bibitem{Kinney}
Kinney, R., Crucitti, P., Albert, R., and Latora, V.
\newblock {\em The European Physical Journal B}{ \bf 46}, 101 (2005).

\bibitem{Mamede}
Mamede, G.~L., Ara{\'u}jo, N.~A., Schneider, C.~M., de~Ara{\'u}jo, J.~C., and
  Herrmann, H.~J.
\newblock {\em Proc Natl Acad Sci USA}{ \bf 109}, 7191--7195 (2012).

\bibitem{Schneider}
Schneider, C.~M., Moreira, A.~A., Andrade~Jr, J.~S., Havlin, S., and Herrmann,
  H.~J.
\newblock {\em Proc Natl Acad Sci USA}{ \bf 108}, 3838--3841 (2011).

\bibitem{Amaral}
Amaral, L. A.~N., Scala, A., Barth{\'e}lemy, M., and Stanley, H.~E.
\newblock {\em Proc Natl Acad Sci USA}{ \bf 97}, 11149--11152 (2000).

\bibitem{Guimera}
Guimer\`{a}, R., Mossa, S., Turtschi, A., and Amaral, L.~N.
\newblock {\em Proc Natl Acad Sci USA}{ \bf 102}, 7794--7799 (2005).

\bibitem{Sales}
Sales-Pardo, M., Guimer\`{a}, R., Moreira, A.~A., and Amaral, L. A.~N.
\newblock {\em Proc Natl Acad Sci USA}{ \bf 104}, 15224--15229 (2007).

\bibitem{Guimera1}
Guimer\`{a}, R., Sales-Pardo, M., and Amaral, L.~A.
\newblock {\em Nature physics}{ \bf 3}, 63--69 (2007).

\bibitem{Cardillo}
Cardillo, A., Zanin, M., G{\'o}mez-Garde{\~n}es, J., Romance, M., del Amo, A.
  J.~G., and Boccaletti, S.
\newblock {\em The European Physical Journal Special Topics}{ \bf 215}, 23--33
  (2013).

\bibitem{Cardillo1}
Cardillo, A., G{\'o}mez-Garde{\~n}es, J., Zanin, M., Romance, M., Papo, D., del
  Pozo, F., and Boccaletti, S.
\newblock {\em Sci Rep}{ \bf 3}, 1344 (2013).

\bibitem{OpenFlights}
Patokallio, J.
\newblock \url{http://openflights.org/}.

\bibitem{Newman2}
Newman, M.~E.
\newblock {\em Networks: An Introduction}.
\newblock Oxford University Press, Inc.,  (2010).

\bibitem{Newman3}
Newman, M.~E.
\newblock {\em Proc Natl Acad Sci USA}{ \bf 103}, 8577--8582 (2006).

\bibitem{Blondel}
Blondel, V.~D., Guillaume, J.-L., Lambiotte, R., and Lefebvre, E.
\newblock {\em Journal of Statistical Mechanics: Theory and Experiment}{ \bf
  2008}, P10008 (2008).

\bibitem{Dorogovtsev1}
Dorogovtsev, S.~N., Goltsev, A.~V., and Mendes, J. F.~F.
\newblock {\em Phys Rev Let}{ \bf 96}, 040601 (2006).

\bibitem{Cohen1}
Cohen, R., Erez, K., Ben-Avraham, D., and Havlin, S.
\newblock {\em Phys Rev Let}{ \bf 86}, 3682 (2001).

\bibitem{Watts1}
Watts, D.~J. and Strogatz, S.~H.
\newblock {\em Nature}{ \bf 393}, 440--442 (1998).

\bibitem{Barrat}
Barrat, A., Barth{\'e}lemy, M., Pastor-Satorras, R., and Vespignani, A.
\newblock {\em Proc Natl Acad Sci USA}{ \bf 101}, 3747--3752 (2004).

\end{thebibliography}
\end{document}